\documentclass[prd,preprintnumbers,superscriptaddress,twocolumn,aps,amsmath,amssymb,showpacs,nofootinbib]{revtex4-2}
\pdfoutput=1
\usepackage{graphicx,array}
\usepackage{hyperref}
\usepackage{color}
\usepackage{amsmath,mathrsfs,amssymb,slashed,latexsym}
\usepackage{bbm}
\usepackage{braket}
\usepackage{placeins}
\usepackage{adjustbox}
\usepackage{subcaption}
\usepackage{enumitem}
\usepackage{comment}
\usepackage[compat=1.1.0]{tikz-feynman}

\def\beq{\begin{equation}}
\def\eeq{\end{equation}}

\newcommand{\tr}{\mathrm{Tr}}
\newcommand{\beT}{b^e_\perp}
\newcommand{\bqT}{b^q_\perp}

\newcommand{\amp}[1]{\mathcal{M}^{#1}_{#1}}

\usepackage{array}
\newcolumntype{P}[1]{>{\centering\arraybackslash}p{#1}}
\newcolumntype{M}[1]{>{\centering\arraybackslash}m{#1}}
\usepackage{tabularray}
\UseTblrLibrary{booktabs}

\definecolor{darkblue}{cmyk}{1,0.4,0,0.3}
\definecolor{violet}{cmyk}{0,1,0,0.2}
\hypersetup{colorlinks, bookmarksnumbered, citecolor=darkblue, linkcolor=darkblue, pdfstartview=FitH, urlcolor=darkblue, linktocpage}

\usepackage[normalem]{ulem}

\begin{document}

\title{Quantum Information at the Electron-Ion Collider}

\author{\large Kun Cheng}\email{kun.cheng@pitt.edu}
\affiliation{Pittsburgh Particle Physics Astrophysics and Cosmology Center,
Department of Physics and Astronomy, University of Pittsburgh, Pittsburgh, PA 15260, USA}
\author{\large Tao Han}\email{than@pitt.edu}
\affiliation{Pittsburgh Particle Physics Astrophysics and Cosmology Center,
Department of Physics and Astronomy, University of Pittsburgh, Pittsburgh, PA 15260, USA}
\author{\large Sokratis Trifinopoulos}\email{trifinos@mit.edu}
\affiliation{Center for Theoretical Physics, Massachusetts Institute of Technology, Cambridge, MA 02139, USA}
\affiliation{Theoretical Physics Department, CERN, Geneva, Switzerland}
\affiliation{Physik-Institut, Universit\"at Z\"urich, 8057 Z\"urich, Switzerland
}%

\preprint{PITT-PACC-2509}
\preprint{CERN-TH-2025-205}
\preprint{MIT-CTP/5943}

\begin{abstract}
We investigate quantum-information-theoretic observables in electron--proton scattering at the Electron-Ion Collider (EIC).  
Our analysis focuses on entanglement and magic, two complementary indicators of non-classicality in quantum states.  
We show that while unpolarized and  longitudinally polarized beams yield unentangled separable outcomes, transverse beam polarization enables the generation of entangled and non-stabilizer states. This result holds for both elastic and deep  inelastic electron-proton scattering in QED. In the deep inelastic regime, the degree of quantum correlation is governed by the transversity parton distribution functions, providing a novel perspective on spin dynamics within QCD.   
These results establish the EIC as a promising environment for generating entangled and non-stabilizer states in high-energy physics, and they highlight opportunities for future lepton–hadron colliders to extend such studies into new kinematic domains.  
\end{abstract}

\maketitle

\section{Introduction}

Quantum information theory (QI) is most commonly studied in low-energy platforms where the basic degrees of freedom are well-controlled qubits or qudits, realized, for example, in trapped ions, superconducting circuits or photonic systems. 
More recently, there has been growing interest in exploring these ideas in collider experiments, since they provide the main controllable terrestrial quantum systems with a large amount of data at the high-energy frontier (see Ref.~\cite{Afik:2025ejh} for a recent review). 
In a collider setting, a natural quantum number that can carry quantum information is the spin of the particles participating in a scattering process. 
Treating each spin as a qubit, the final state of a scattering process can be viewed as a multi-qubit quantum state, allowing a systematic study of QI-theoretic properties that arise directly from the fundamental interactions of the standard model (SM). 

Two key quantities that capture complementary aspects of non-classicality are quantum \emph{entanglement} and \emph{magic}. 
Entanglement characterizes the non-separability of a quantum state, distinguishing it from any classical mixture of product states. 
Magic, or non-stabilizerness, measures the departure of a state from the set of stabilizer states generated by Clifford gates, which can be efficiently simulated on classical computers~\cite{Nielsen:2012yss}. 
States with nonzero magic therefore constitute a genuine computational resource, essential for universal fault-tolerant quantum computation. 
Understanding how entanglement and magic are produced in high-energy scattering processes is the target of current explorations: the entangling properties of SM interactions have been studied from a theoretical perspective~\cite{Balasubramanian:2011wt,Seki:2014cgq,Peschanski:2016hgk,Grignani:2016igg,Kharzeev:2017qzs,Fan:2017hcd,Fan:2017mth,Cervera-Lierta:2017tdt,Beane:2018oxh,Rigobello:2021fxw,Low:2021ufv,Liu:2022grf,Fedida:2022izl,Cheung:2023hkq,Carena:2023vjc,Aoude:2024xpx,Low:2024mrk,Low:2024hvn,Thaler:2024anb,McGinnis:2025brt,Carena:2025wyh,Liu:2025pny,Hu:2025lua,Sou:2025tyf}, 
but also experimentally probed at the LHC, with a canonical example being the production of entangled spin states of top quark pairs~\cite{Afik:2020onf,Dong:2023xiw,Aoude:2022imd,Fabbrichesi:2021npl,Severi:2021cnj,Afik:2022kwm,Aguilar-Saavedra:2022uye,Fabbrichesi:2022ovb,Afik:2022dgh,Severi:2022qjy,Aguilar-Saavedra:2023hss,Han:2023fci,Cheng:2023qmz,Cheng:2024btk,Simpson:2024hbr,Aguilar-Saavedra:2024hwd,ATLAS:2023fsd,CMS:2024pts,Barr:2024djo,Low:2025aqq}, as well as in future colliders~\cite{Altakach:2022ywa,Bi:2023uop,Aoude:2023hxv,Ma:2023yvd,Fabbrichesi:2023cev,Ruzi:2024cbt,Maltoni:2024csn,Ding:2025mzj,Qi:2025onf}.
In contrast, the study of magic in fundamental interactions has only just begun, with first results emerging very recently~\cite{White:2024nuc,Robin:2024bdz,Aoude:2025jzc,Fabbrichesi:2025ywl,Busoni:2025dns,Liu:2025qfl,Gargalionis:2025iqs,Liu:2025bgw,CMS:2025cim}.

An additional opportunity arises when the ability to prepare the initial-state spin is present. 
While at the LHC the colliding proton beams are not polarized, this feature is envisaged in future colliders such as the Electron-Ion Collider (EIC)~\cite{Accardi:2012qut,AbdulKhalek:2021gbh} at Brookhaven National Laboratory. 
The EIC will collide polarized electron beams of up to $18$~GeV with polarized protons or ions of up to $275$~GeV per nucleon, with a projected integrated luminosity of order $100~\text{fb}^{-1}$. 
Both the electron and proton beams can be prepared with a varying degree of polarization, allowing control over the initial spin state of the system. 
This makes it possible to initialize the scattering process in well-defined quantum states, so that the amount of entanglement or magic produced is determined by the chosen state preparation and the structure of the SM interactions. 

In this work we investigate how entanglement and magic can be generated and characterized in electron--proton scattering at the EIC. 
We focus on the case where both beams are transversely polarized,\footnote{The scattering with a transversely polarized nucleon target or beam has been considered at RHIC~\cite{Bunce:2000uv}, HERMES~\cite{HERMES:2004mhh}, COMPASS~\cite{COMPASS:2005csq} and the  EIC~\cite{AbdulKhalek:2021gbh}. } 
corresponding to coherent superpositions of helicity eigenstates.
This choice is motivated by the fact that longitudinally polarized or unpolarized beams only lead to classical correlation in the final state, as we demonstrate explicitly below. 
From the scattering amplitudes one finds that maximal entanglement occurs in kinematic regions where elastic scattering is not realized at EIC energies. 
Consequently, the relevant setting is deep inelastic scattering, where the outgoing state includes hadronic fragments and the entangled subsystem is formed by the electron and the partonic quark. 
In this regime, the degree of entanglement is controlled by the transversity parton distribution functions, making transverse polarization at the EIC an important asset for probing quantum information in Standard Model processes. 

This Letter is organized as follows.  
In Sec.~\ref{sec:essentials} we briefly review the QI concepts relevant for our study, namely entanglement and magic.  
Sec.~\ref{sec:pol_ep} presents the framework for polarized $ep$ scattering, including the treatment of spin density matrices and scattering amplitudes.  
In Sec.~\ref{sec:DIS} we perform the analysis of deep inelastic scattering, where partonic degrees of freedom enter through transversity distributions.  
Finally, in Sec.~\ref{sec:conclusions} we summarize our main findings and comment on future directions for quantum information studies at the EIC and beyond.

\section{Quantum Essentials}
\label{sec:essentials}

\noindent 
\textbf{Quantum Entanglement.} 
A bipartite quantum state $\rho$ is called separable if it can be expressed as a convex combination of product states~\cite{PhysRevLett.77.1413,HORODECKI1997333}:  
\begin{equation}\label{eq:separable}
    \rho=\sum_n p_n\, \rho^A_n \otimes \rho^B_n, 
    \qquad \sum_n p_n=1, \quad p_n\geq0,
\end{equation}
where $\rho^{A,B}_n$ are the states in the two subsystems $A$ and $B$.  
If a state cannot be written in the form of Eq.~\eqref{eq:separable}, it is entangled.  

For two-qubit systems, a widely used measure of entanglement is the \emph{concurrence}~\cite{PhysRevLett.80.2245}:  
\begin{equation}
    \mathscr{C}[\rho] \equiv \max\!\left(0, \lambda_1-\lambda_2-\lambda_3-\lambda_4\right), 
    \ \  0\leq \mathscr{C}[\rho]\leq 1,
    \label{eq:Concurrence}
\end{equation}
where $\{\lambda_i\}$ are the eigenvalues in decreasing order of the matrix 
$\sqrt{\sqrt{\rho}\,\tilde{\rho}\,\sqrt{\rho}}$, with 
$\tilde{\rho}=(\sigma_2\otimes\sigma_2)\rho^*(\sigma_2\otimes\sigma_2)$ and $\rho^*$ the complex conjugate of $\rho$.  
Concurrence vanishes for separable states, is positive for entangled states, and reaches unity for maximally entangled Bell states.  

\noindent 
\textbf{Quantum Magic.}  
To formalize the notion of non-stabilizerness, it is useful to recall the algebraic structure underlying qubit systems.  
For an $n$-qubit Hilbert space, one defines the set of \emph{Pauli strings}  
\begin{equation}
  {\cal P}_n = P_1 \otimes P_2 \otimes \ldots \otimes P_n , 
  \qquad P_i \in \{\mathbbm{1},\sigma_1,\sigma_2,\sigma_3\}\,,
  \label{Paulistring}
\end{equation}
which generates the Pauli group.  
The subgroup of unitaries that map Pauli strings to each other in conjugation forms the \emph{Clifford group} $\mathcal{C}_n$.  
Acting with Clifford gates on the reference state $|0\cdots 0\rangle$ produces the set of \emph{stabilizer states}, which admit efficient classical simulation by the Gottesman--Knill theorem.  

A convenient way to diagnose whether a general state $\rho$ is stabilizer or not is through its \emph{Pauli spectrum}, the collection of expectation values $\{\mathrm{Tr}[P\rho], \, P\in{\cal P}_n\}$.  
To quantify deviations from the stabilizer set, one may introduce the \emph{stabilizer Rényi entropies}~\cite{Leone:2021rzd}:  
\begin{equation}
  M_k(\rho)=-\frac{1}{1-k}\log_2\!\left(\frac{\sum_{P\in{\cal P}_n} (\mathrm{Tr}[P\rho])^{2k}}
  {\sum_{P\in{\cal P}_n} (\mathrm{Tr}[P\rho])^2}\right),
  \label{Mqdef}
\end{equation}
defined for integer $k\geq 2$.  
These measures vanish for stabilizer states, increase as the state departs from the Clifford subspace, and are also additive when combining quantum systems.
In practice, the $k=2$ case, known as the \emph{second stabilizer Rényi entropy} (SSRE), is widely used as a robust probe of magic. Magic isolates the part of the quantum structure that cannot arise from Clifford circuits and stabilizer measurements alone~\cite{Nielsen:2012yss}. It is a necessary resource for universal and fault-tolerant quantum computation through state injection or other non-Clifford operations. A nonzero SSRE  entropy signals failure of efficient classical simulation by stabilizer methods, and thus reveals computational power in the state beyond what entanglement alone certifies.

\section{Polarized $ep$ Scattering}
\label{sec:pol_ep}

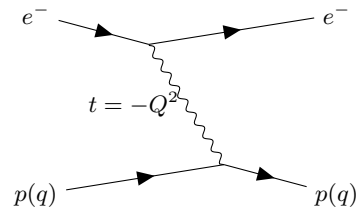
\begin{figure}
    \centering
        \begin{tikzpicture}[baseline=(center.base)]
        \begin{feynman}
            \vertex (i1) at (-2.0cm,1.2cm){$e^-$};
            \vertex (i2) at (-2.0cm,-1.2cm){$p(q)$};
            \vertex (v1) at (-0.5cm,0.8cm);
            \vertex (v2) at (0.5cm,-0.8cm);
            \vertex (o1) at (2.0cm,1.2cm){$e^-$};
            \vertex (o2) at (2.0cm,-1.2cm){$p(q)$};
            \vertex (center) at (0,0){};
            \vertex (label) at (-0.7cm,0){$t=-Q^2$};
            \diagram{(i1)--[fermion](v1)--[fermion](o1),(i2)--[fermion](v2)--[fermion](o2),(v1)--[boson](v2)};
        \end{feynman}
    \end{tikzpicture}
    \caption{Elastic $e^-p$ ($e^- q$) scattering at the EIC.}
    \label{fig:epep}
\end{figure}

\noindent 
\textbf{Initial and final state density matrices.} In the following, we focus on the case where both electron and proton initial beams are transversely polarized along the $x$ direction, with their spin density matrices given by
\begin{equation}\label{eq:rhoOfInitialBeam}
    \rho^{e/p} = \frac{I_2 + \vec b_\perp^{e/p}\cdot \vec\sigma_\perp}{2},
\end{equation}
respectively, where $I_2$ is the 2-dimensional identity matrix, $\vec\sigma_\perp=(\sigma_1,\sigma_2)$ is the Pauli matrix in the transverse plane,  $\vec b_\perp^{e/p}$ is the transverse polarization vector of the electron or proton beam.
For simplicity, we consider the case that the electron and proton have the same direction of transverse polarization, which we choose as the $x$-direction corresponding to $\sigma_1$.
A transversely polarized electron/proton state is a coherent superposition of the spin state with positive and negative helicity.
In the case of $|\vec b_\perp|=1$, the state corresponds to the pure state $(\ket{\uparrow}+\ket{\downarrow})/\sqrt{2}$. 

When the initial state is a mixed state, generally the final spin state is also a mixed state.  Up to a normalization factor, the spin density matrix of the final state pair $ep$ (or the pair $eq$ for DIS) is
\begin{equation}
    \rho_{\alpha\bar\alpha,\alpha'\bar\alpha'} \propto \sum_{\lambda\bar\lambda\lambda'\bar\lambda'}\mathcal{M}_{\alpha\bar\alpha}^{\lambda\bar\lambda}(\mathbf{k})\rho^{e}_{\lambda\lambda'} \rho^{p/q}_{\bar{\lambda}\bar{\lambda}'}(\mathcal{M}_{\alpha'\bar\alpha'}^{\lambda'\bar\lambda'}(\mathbf{k}))^\dagger,
    \label{eq:mixed_state}
\end{equation}
where $\mathcal{M}$ denote the perturbative scattering amplitudes. Here, the initial electron and proton beams are prepared separately with no correlation, so that their density matrix is in separable form.

\noindent 
\textbf{Scattering Amplitudes.}

We start from the elastic $e^-p$ scattering, but the formalism presented in this section is equally applicable to the  $e^-q$ scattering when considering the deep inelastic process. When the proton is boosted to high energies ($\sqrt{s}\gg m_p$), vector interactions guarantee that the outgoing proton (electron) has the same helicity as the incoming proton (electron).  Then via $t$-channel photon exchange, there are only four non-zero amplitudes $\mathcal{M}^{\alpha\beta}_{\alpha'\beta'}\equiv \mathcal{M}(e^-(\alpha) p (\beta) \to e^-(\alpha') p(\beta') ) $ in the helicity basis: 
\begin{subequations} \label{eq:amplitudes}
\begin{align}
    \amp{\uparrow\Uparrow}& \approx 2 Q_p e^2 \frac{s+t}{t} e^{i\phi}\,, \\
    \amp{\uparrow\Downarrow} = \amp{\downarrow\Uparrow} &\approx 2 Q_p  e^2 \frac{s}{t}\,,
    \\
    \amp{\downarrow\Downarrow}& \approx  2 Q_p e^2 \frac{s+t}{t} e^{-i\phi} , 
\end{align}
\end{subequations}
where $Q_p=+1$ is the proton charge and $\phi$ the azimuthal angle of the outgoing electron scattering plane with respect to a transverse direction to be specified later. 
The amplitudes with massive proton and form factors are given in Appendix~\ref{app:helicityAMP}.
Here, the upper indices $\uparrow\downarrow$ and $\Uparrow\Downarrow$ denote the spin of the initial electron and proton, both quantized along the direction of the electron beam. The lower indices  $\uparrow\downarrow$ and $\Uparrow\Downarrow$  denote the spin of the outgoing electron and proton, both quantized along the direction of the outgoing electron.

We observe that the transition matrix is diagonal. Consequently, with the left- or right-handed polarized initial state $e^-$, the spin state $\ket{ep}$ of the outgoing state pair $e^-p$ is in a separable form 
\begin{equation}\label{eq:LeftepRightep}
    \ket{\uparrow} \otimes\ket{p} \text{~~or~~} \ket{\downarrow}\otimes \ket{p'} , 
\end{equation}
where the helicity state of the outgoing $e^-$ is the same as the initial state.
Similarly, if the initial electron beam is unpolarized, i.e., an incoherent mixture of left- and right-handed polarized states, then the final state $e^- p$ is a mixture of the states in Eq.~\eqref{eq:LeftepRightep}, which is still a separable state that can be written in the form of Eq.~\eqref{eq:separable}.

\begin{figure}
    \centering
    \includegraphics[scale=1]{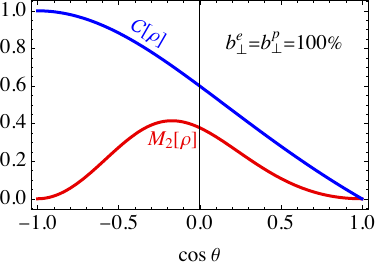}
    \caption{The concurrence and stabilizer Rényi entropy as a function of scattering angle for elastic $ep$ scattering with 100\% transversely polarized beams.}
    \label{fig:C_M2_purestate}
\end{figure}

Although the $t$-channel scattering of unpolarized or longitudinally polarized beams only produces separable final states,  quantum entanglement can arise with a transverse polarization.
If the initial state is a coherent superposition of different helicity states, then the final state is also a coherent superposition of the two states in Eq.~\eqref{eq:LeftepRightep} and could be entangled.
For example, if both $e^-$ and $p$ beams are pure states given by $\frac{1}{\sqrt{2}}(\ket{\uparrow}+\ket{\downarrow})$, from the amplitudes in Eq.~\eqref{eq:amplitudes}, the final state vector is
\begin{align}
    \ket{ep}\propto &
    \left(\frac{s-Q^2}{s}\right)
    \left(e^{i\phi}\ket{\uparrow\uparrow}+e^{-i\phi}\ket{\downarrow\downarrow}\right)+\ket{\uparrow\downarrow} + \ket{\downarrow\uparrow}\,,
\end{align}
where $Q^2=-t$ is the momentum transfer in the scattering, and $\phi$ is defined with respect to the transverse polarization of the initial beams.
The concurrence and magic of the final state produced from 100\% transversely polarized initial states are shown in Fig.~\ref{fig:C_M2_purestate}.
In the forward elastic scattering region when $Q^2\to 0$, the final state $ep$ pair is separable because the spin configuration of the initial state is left unchanged after the scattering. In the backward scattering region with maximal momentum transfer $Q^2 \approx s$, the spin state of $ep$ is a maximally entangled spin triplet.

\section{Deep Inelastic Scattering}
\label{sec:DIS}

The scattering between electron and proton introduces sizable quantum entanglement only when the momentum transfer $Q^2$ is comparable to $s$. However, in the high-energy limit with $Q^2\gg m_p^2$, the elastic scattering is negligible and the deep inelastic scattering (DIS) becomes the dominant process.
Therefore, we next focus on the parton level $e^-q\to e^-q$ scattering and the helicity amplitudes are obtained by replacing the proton charge $Q_p$ with the respective quark charges $Q_q$ in Eq.~\eqref{eq:amplitudes}. The outgoing quark $q$ hadronizes into a jet.

\begin{figure}
    \centering
    \includegraphics[scale=0.9]{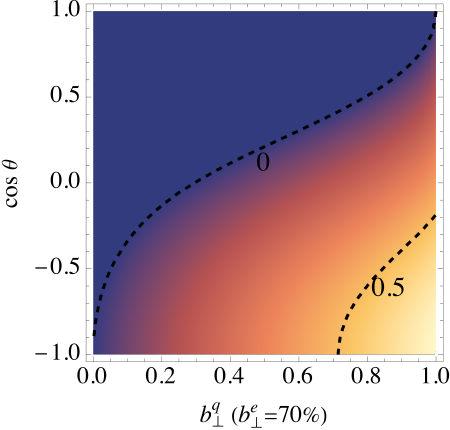}
    \caption{Concurrence of final state $eq$ pair as a function of parton level scattering angle and the transverse polarization of the quark.  The transverse polarization of the electron beam is fixed to be 70\%.}
    \label{fig:BxCosConcurrence}
\end{figure}

Consider the realistic case where the initial states are partially transversely polarized, the spin density matrix of the final state $e^-q$ pair from Eq.~\eqref{eq:mixed_state} in the limit of backward scattering $(\cos\theta=-1)$ is given by
\begin{equation}\label{eq:backwardScattering}
    \rho_{\rm bk}=\frac{I_4+C_{ij}\sigma_i\otimes\sigma_j}{4}, \quad 
    C_{ij}=\begin{pmatrix}
        b^e_\perp b^q_\perp & 0 &0 \\
        0 & b^e_\perp b^q_\perp  &0 \\
        0 & 0 & -1 \\
    \end{pmatrix}
\end{equation}
where $b^{e/q}_\perp$ is the degree of transverse polarization of the initial electron/quark state defined in Eq.~\eqref{eq:rhoOfInitialBeam}, and the concurrence is simply given by
\begin{equation}
    \mathscr{C}= b^e_\perp b^q_\perp.
\end{equation}
For general scattering angles, the spin density matrix of the final state $eq$ pair is given in Eq.~\eqref{eq:CijGeneralTheta}, and the concurrence as a function of initial beam polarization and scattering angle is shown in Fig.~\ref{fig:BxCosConcurrence}, where we fix the transverse polarization of the electron beam to be $70\%$ and vary the polarization of the initial quark.

We see that the entanglement of the final state crucially maximizes in back-scattering region, and increases with the degree of transverse polarization of the initial state. The dependence of the entanglement on the polarization gives us the opportunity to control the state production by changing the initial state polarization.
Although the polarizations of initial electron and proton beams are usually fixed during the collision, as we introduce below, the transverse polarization of quark from a proton beam with a fixed transverse polarization can vary with the kinematics $x$ and $Q^2$. 
Therefore, we are able to choose different quark polarizations from the kinematics.

The transverse polarization of quarks inside a proton is described by the transversity parton distribution functions (transversity PDFs) $h_1(x)$~\cite{Ralston:1979ys,Artru:1989zv,Jaffe:1991kp,Jaffe:1991ra,Cortes:1991ja}, which is the number density of quarks polarized in a transverse direction $+\hat{n}$ with a given longitudinal momentum fraction $x$ minus the number density of quarks polarized in the opposite direction $-\hat{n}$, when the hadron spin points in the direction $+\hat{n}$.
Therefore, the transverse polarization vector $\vec{b}^q_{\perp}$ of the quark from a transversely polarized proton with polarization vector $\vec{b}^p_{\perp}$ is $\vec{b}^q_{\perp}= \frac{h_{1,q}(x)}{f_q(x)}\vec{b}^p_{\perp} $ where $f_q(x)$ is the unpolarized PDF.
Since the light quarks are indistinguishable, we should treat the spin state of quarks as a mixed state including all the flavors. Effectively, the average transverse polarization of quarks that participate in DIS is
\begin{equation}\label{eq:PqPp}
    \vec{b}^{\langle q\rangle}_\perp =\alpha  \vec{b}^p_{\perp},\quad \alpha\equiv  \frac{\sum_q e_q^2 h_{1,q}(x)}{\sum_q e_q^2 f_q(x)}.
\end{equation}
The second-generation quarks contribute only to the unpolarized PDF and do not contribute to the transversity PDF. 

\begin{figure}
    \centering
    \includegraphics[scale=1]{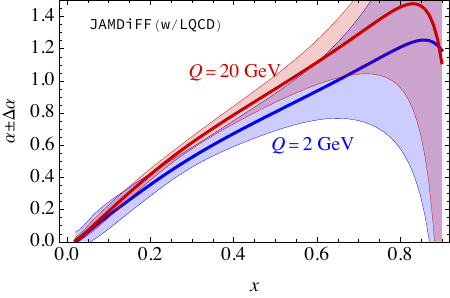}
    \caption{Center value and uncertainty of the effective transverse polarization fraction of the proton carried by initial state quarks.}
    \label{fig:factorAlpha}
\end{figure}

We show the center value and the uncertainty of the factor $\alpha$ in Fig.~\ref{fig:factorAlpha} with the transversity PDF from JAMDiFF(w/LQCD)~\cite{Cocuzza:2023vqs}. Given the transverse polarization of the proton beam, we see that overall the transverse polarization of the quarks increases with the momentum fraction $x$.
However, the uncertainty $\Delta\alpha$ diverges in large $x$ because both $h_1(x)$ and $f(x)$ approach zero. Due to the large uncertainty, the effective polarization fraction $\alpha$ easily violates the unitarity bound $\alpha<1$ when $x\to 1$. Therefore, as a conservative estimate, we use the  $1\sigma$ lower limit of $h_1(x)$ to calculate the factor $\alpha$, and cut the region where the uncertainty is too large $\Delta h_1(x)>|h_1(x)|$. 

\begin{figure}
    \centering
    \includegraphics[scale=0.9]{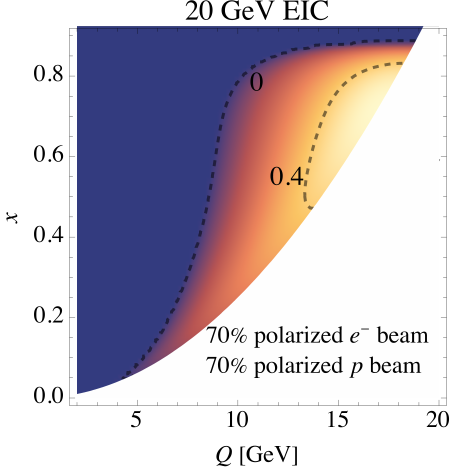}
    \caption{Concurrence of $eq$ pair from DIS process at a $\sqrt{s} =20\,\rm GeV$ EIC.  The kinematic limit $Q^2=x(20~{\rm GeV})^2$ corresponds to backward scattering with $\cos\theta=-1$.
    }
    \label{fig:QxConcurrence}
\end{figure}

With Eq.~\eqref{eq:PqPp},
the dependence of concurrence on the transverse polarization of the initial quark in Fig.~\ref{fig:BxCosConcurrence} can be translated into the dependence on the DIS kinematics $Q^2$ and $x$. In Fig.~\ref{fig:QxConcurrence}, we show the concurrence of the $eq$ pair produced from DIS at a 20 GeV EIC. 

The upper limit of $Q^2$ corresponds to the backward scattering where a largely entangled pair $eq$ can be produced if the initial states are transversely polarized. As we can see from Fig.~\ref{fig:factorAlpha}, the transverse polarization of the initial quarks has a mild dependence on $Q^2$ and is basically determined by $x$. When $x$ is around $0.6\sim0.8$, the transverse polarization of the initial quark maximizes, and so does the concurrence of the final state.

\section{Discussions and Conclusions}
\label{sec:conclusions}

In this work, we have explored how QI properties emerge in polarized electron--proton scattering at the EIC.  
We demonstrated that nontrivial quantum correlations arise only when both beams are transversely polarized, leading to final states with measurable entanglement and magic.  
Since the elastic kinematics is not favorable at higher energies, deep inelastic scattering provides the relevant environment, where the strength of quantum correlations is governed by the transversity parton distribution functions and reaches its maximum in the backward-scattering region.  

Unlike other collider measurements that typically infer the spins of heavy particles through decay distributions, illustrated for the EIC in the DIS process  \cite{Qi:2025onf}, we formulated the entanglement observables of the final state $e^-q$ pair (with the quark hadronizing into a jet)  by identifying the initial state polarization from the DIS scattering kinematics. This is similar to the kinematic approach described in Ref.~\cite{Cheng:2024rxi}, but additional polarization information needs to be constructed from scattering kinematics with the transversity PDF.
For a complementary approach without relying on the transversity PDF, one needs to measure the quark spin from fragmentation~\cite{Cheng:2025cuv} and electron spin with future devices.

We note that at low momentum transfer $Q \sim$1 GeV, one may include a momentum-dependent form factor, as given in Eq.~(\ref{eq:form}). This would only slightly modify the numerical results, but would not change the qualitative behavior. There are also inelastic contributions in low momentum transfer with multiple hadrons in the final state \cite{Blumlein:2002fw,Sofiatti:2011yi,Guzey:2023syh}. However, in this region, a two-qubit final state is no longer well defined. We thus focused on the elastic and the DIS processes.

Although we have limited our analysis to the initial low energy of the EIC $\sqrt s=20$ GeV, we expect our conclusions to hold at higher EIC energies near 100 GeV, namely, that the entanglement is present only with beam transverse polarization and reaches the maximum in the back-scattering region. At even higher energies above the electroweak scale, the contribution to $ep$ collisions from the neutral-current $Z$ exchange becomes significant, and chiral couplings modify the helicity amplitudes in  Eq.~\eqref{eq:amplitudes} accordingly.

Looking to the future, one avenue of progress would be the development of techniques for direct spin measurements of stable particles at colliders. 
This would bring high-energy experiments closer to low-energy quantum platforms where spin readout is standard. 
Emerging ideas, such as the use of nitrogen-vacancy centers in diamond~\cite{Doser:2022knm}, hint at possible paths forward, although their realization in collider environments remains speculative. 
Although challenging on EIC timescales, such technologies could become natural ingredients for detectors of next-generation lepton–ion colliders at higher energies, such as the TeV-scale Muon synchronization Ion Collider (MuSIC)~\cite{Acosta:2021qpx,Acosta:2022ejc,Davoudiasl:2024fiz}, where new kinematic regimes of quantum tomography can be probed.

\begin{acknowledgments}

The authors  thank Yoav Afik and Keping Xie for useful discussions.
This work was supported in part by the US Department of Energy under grant No. DE-SC0007914 and in part by Pitt PACC.
ST was supported by the Office of High Energy Physics of the US Department of Energy (DOE) under Grant No.~DE-SC0012567, and by the DOE QuantISED program through the theory consortium “Intersections of QIS and Theoretical Particle Physics” at Fermilab (FNAL 20-17).
ST is also supported by the Swiss National Science Foundation project number PZ00P2\_223581, and acknowledges the CERN TH Department for hospitality while this research was being carried out.

\end{acknowledgments}

\appendix
\clearpage
\begin{widetext}
\section{Massive Helicity Amplitudes}\label{app:helicityAMP}

In the main text, we focus on the high-energy limit.
Here, we also lay out the general helicity amplitudes with massive protons, and show that the low-energy elastic scattering usually produces classical correlation.
We work in the c.m. frame of initial $ep$ beams and the momenta are
\begin{equation}
    p_e = (E_e, 0,0, E_e),~~~~
    p_p = (E_p, 0,0, -E_e)
\end{equation}
where we have $E_p^2 - E_e^2 = m_p^2$.  The outgoing electron and proton momenta are
\begin{equation}
    p_e'=E_e(1,\sin\theta\cos\phi,\sin\theta\sin\phi,\cos\theta),\qquad
    p_p'=E_p(1, \beta\sin\theta \cos\phi,\beta\sin\theta\sin\phi,\beta\cos\theta), \quad \beta = \frac{E_e}{E_p},
\end{equation}
where $\beta$ is the velocity of the proton.
We keep the azimuthal angle $\phi$ explicit because the scattering process would have a non-trivial $\phi$ dependence in the presence of transverse polarization of initial beams. For simplicity, throughout this work, we consider the case that both initial beams are transversely polarized along the same direction, and the azimuthal angle $\phi$ is defined with respect to this direction.
The Mandelstam variable $s$ and $t$, are
\begin{align}
    s = (E_e + E_p)^2,\qquad 
    t = -2E_e^2 (1- \cos\theta). 
\end{align}

The elastic scattering amplitude is given by $\mathcal{M}=\frac{e^2}{t}J_{e,\mu}J_{p}^\mu$ with
\begin{equation}
    J^\mu_{e,\alpha\alpha'} = \bar u_{\alpha'}(p_e') \gamma^\mu u_{\alpha}(p_e),\qquad
    J^\mu_{p,\beta\beta'} = \bar u_{\beta'}(p_p') \Gamma^\mu u_{\beta}(p_p) , 
\end{equation}
where $(\alpha,\alpha')$ are the spin indices of the incoming and outgoing electron, while $(\beta,\beta')$ are the spin indices of the proton. The proton vertex with form factors is
\begin{equation}
    \Gamma^\mu=\gamma^\mu F_1(q^2)+\frac{i\sigma^{\mu\nu}q_\nu}{2m_p}F_2(q^2), \qquad  q=p_p'-p_p,~ \sigma^{\mu\nu}=\frac{i}{2}[\gamma^\mu,\gamma^\nu]. 
    \label{eq:form}
\end{equation}
The helicity of the electron does not change in the $t$-channel scattering but the helicity of the proton could change due to its mass.  The transition matrix is
\begin{align}
\renewcommand{\arraystretch}{1.8}
    \mathcal{T}&=\begin{pmatrix}
        \mathcal{M}^{\uparrow\Uparrow}_{\uparrow\Uparrow} &  \mathcal{M}^{\uparrow\Downarrow}_{\uparrow\Uparrow} &  0 & 0 \\
        \mathcal{M}^{\uparrow\Uparrow}_{\uparrow\Downarrow} &  \mathcal{M}^{\uparrow\Downarrow}_{\uparrow\Downarrow} &  0 & 0 \\ 
        0&0 &\mathcal{M}^{\downarrow\Uparrow}_{\downarrow\Uparrow} & \mathcal{M}^{\downarrow\Downarrow}_{\downarrow\Uparrow} \\
        0&0 &\mathcal{M}^{\downarrow\Uparrow}_{\downarrow\Downarrow} & \mathcal{M}^{\downarrow\Downarrow}_{\downarrow\Downarrow}
    \end{pmatrix} \nonumber \\
    &= 2 e^2 \frac{E_e E_p}{t} F_1(q^2)
    \begin{pmatrix}
        e^{i\phi} (1+\beta)(1+c_\theta) &  s_\theta \sqrt{1-\beta^2} &  0 & 0 \\
        -e^{i\phi}s_\theta \sqrt{1-\beta^2} &  1+3\beta+(1-\beta)c_\theta &  0 & 0 \\ 
        0&0 &1+3\beta+(1-\beta)c_\theta & e^{-i\phi}s_\theta \sqrt{1-\beta^2} \\
        0&0 &s_\theta \sqrt{1-\beta^2} &  e^{-i\phi} (1+\beta)(1+c_\theta)
    \end{pmatrix} \nonumber\\
    &~~+4e^2\frac{E_e E_p}{t}F_2(q^2) \begin{pmatrix}
        0 & - s_\theta\frac{\beta(\beta+1)}{\sqrt{1-\beta^2}} & 0 & 0 \\
        e^{i\phi}s_\theta\frac{\beta(\beta+1)}{\sqrt{1-\beta^2}} &  2\beta(1-c_\theta) & 0 & 0 \\
        0 &  0 &   2\beta(1-c_\theta) & -e^{-i\phi}s_\theta\frac{\beta(\beta+1)}{\sqrt{1-\beta^2}} \\
        0 & 0  & s_\theta\frac{\beta(\beta+1)}{\sqrt{1-\beta^2}} & 0 \\
    \end{pmatrix}
\end{align}
The spin of both the initial electron and proton are quantized along the electron beam direction; the spin of both final states are defined along the electron outgoing direction in the scattering c.m. frame.

We next confirm that the entanglement between the final state still vanishes when $Q^2\to 0$.
In the low-energy limit when elastic scattering dominates and $\beta\to0$, the transition matrix is
\begin{align}
\renewcommand{\arraystretch}{1.8}
    \mathcal{T}& \propto
    \begin{pmatrix}
        e^{i\phi} (1+c_\theta) &  s_\theta  &  0 & 0 \\
        -e^{i\phi}s_\theta  &  1+c_\theta &  0 & 0 \\ 
        0&0 &1+c_\theta & e^{-i\phi}s_\theta  \\
        0&0 &s_\theta  &  e^{-i\phi} (1+c_\theta) 
    \end{pmatrix} . 
\end{align}
Multiplying the transition matrix by the spin state vector $\sim(\ket{\uparrow}+\ket{\downarrow})\otimes(\ket{\uparrow}+\ket{\downarrow})$, we see that even when both the initial electron and proton are transversely polarized, the final state is still in a separable form,
\begin{equation}
    \ket{ep}_{\rm final} \propto (\ket{\uparrow}+e^{-i\phi}\ket{\downarrow})\otimes\Big( ((1+c_\theta)e^{i\phi}+s_\theta)\ket{\uparrow}+(1+c_\theta-e^{i\phi}s_\theta)\ket{\downarrow}\Big). 
\end{equation}
At higher scattering energy when $\beta\neq 0$, the transition matrix is diagonal when $\cos\theta\to 1$. When both initial beams are transversely polarized, the final state is still in a separable form
\begin{equation}
    \ket{ep}_{\rm final} \propto e^{i\phi}\ket{\uparrow\uparrow}+e^{-i\phi}\ket{\downarrow\downarrow}+\ket{\uparrow\downarrow} + \ket{\downarrow\uparrow}.
\end{equation}
Overall, there is no entanglement between final states when $Q^2\to 0$, no matter how we polarize the initial beam.

In Eq.~\eqref{eq:backwardScattering}, we showed the spin density matrix of the $eq$ pair produced in the backward scattering process as a function of initial state polarization.  Here, we also present the spin density matrix of the $eq$ pair at a general scattering angle, which is given by
\begin{equation}
    \rho=\frac{I_4+B^e_i \sigma_i\otimes I_2 + B^q_i I_2\otimes \sigma_i + C_{ij}\sigma_i\otimes \sigma_j}{4}
\end{equation}

\renewcommand{\arraystretch}{2}
\begin{equation}\label{eq:CijGeneralTheta}
\begin{aligned}
    B^e &= \left( \frac{4 \beT(1+c_\theta)}{4+(1+c_\theta)^2} c_\phi,~~ -\frac{4 \beT(1+c_\theta)}{4+(1+c_\theta)^2} s_\phi ,~~ 0\right)  \\
    B^q &= \left( \frac{4 \bqT(1+c_\theta)}{4+(1+c_\theta)^2} c_\phi,~~ -\frac{4 \bqT(1+c_\theta)}{4+(1+c_\theta)^2} s_\phi ,~~ 0\right)  \\
    C_{ij} &= \begin{pmatrix}
        \beT \bqT\frac{(4+(1+c_\theta)^2c_{2\phi})}{4+(1+c_\theta)^2} & -\beT \bqT\frac{(1+c_\theta)^2s_{2\phi}}{4+(1+c_\theta)^2} & 0 \\
        -\beT \bqT \frac{(1+c_\theta)^2s_{2\phi}}{4+(1+c_\theta)^2} & \beT \bqT\frac{(4-(1+c_\theta)^2c_{2\phi})}{4+(1+c_\theta)^2} & 0 \\
        0 & 0 & -\frac{(1-c_\theta)(3+c_\theta)}{4+(1+c_\theta)^2} \\
    \end{pmatrix} .
\end{aligned}
\end{equation}
\end{widetext}

\bibliographystyle{apsrev4-1}
\bibliography{ref}
\end{document}